# Deep Hype in Artificial General Intelligence: Uncertainty, Sociotechnical Fictions and the Governance of AI Futures

*Andreu Belsunces Gonçalves*
*Universitat Oberta de Catalunya - Communication Network and Social Change / Tecnopolítica*

## Abstract

Artificial General Intelligence (AGI) has emerged as one of the most ambitious frontiers of the contemporary technology sector. Promoted by tech leaders and investors, AGI is imagined as a system capable of performing all human intellectual tasks, and potentially exceeding them. Despite lacking clear definition or technical feasibility, AGI has attracted unprecedented levels of investment and political attention, driven by inflated promises of civilisational transformation.

Based on a series of statements and documents from AGI leaders, this paper examines how a multidimensional network of uncertainties sustains what is defined as deep hype: a long-term, overpromissory dynamic that constructs visions of civilisational transformation through a network of uncertainties extending into an undefined future, making its promises nearly impossible to verify in the present, while maintaining attention, investment, and belief. These uncertainties are articulated through sociotechnical fictions: forms of mediated imagination produced within science and technology but not recognised as such since they are sheltered with technoscientific legitimacy. These fictions make not-yet-existing technologies intelligible and desirable, generating surprise, urgency, and expectation.

The analysis shows how AGI deep hype is fuelled by the recursive interaction between uncertainty, fiction, and venture capital speculation. This entanglement plays a key role in advancing a cyberlibertarian and longtermist programme that displaces democratic oversight, frames regulation as obsolete, and positions private actors as the rightful stewards of humanity's technological destiny. By observing the interplay of these elements, this theoretical paper contributes to the field of hype studies and offers critical insight into the governance of technological futures.

# Introduction

Artificial General Intelligence (AGI) is presented by the US technology industry as one of the  most ambitious and speculative horizons of the artificial intelligence (AI) sector. Unlike narrow AI systems designed for specific tasks, AGI aims to become a form of intelligence capable of performing any intellectual function a human can achieve, and potentially surpassing it (REF). In the last few years, AGI has become a matter of hype in technology, investment, media outlets and regulatory circles.

However, the very notion of AGI is controversial (Summerfield, 2023; Morris et al, 2023; Blili-Hamelin et al, 2024; Mitchell, 2024; Ahmed et al, 2024) as well as its technical feasibility is subjected to a strong controversy (Landgrebe and Smith, 2019; Fjelland, 2020; Sublime, 2023; Jaeger, 2023; Schlereth, 2025). This, nevertheless, did not prevent OpenAI, the company that mainstreamed this not-yet-existing technology, to raise the largest single funding round ever by a private tech startup.

Since its conception in the early twenty-first century, the promise of AGI has been championed by cyberlibertarians and longtermists: technologists, investors, and writers who reject democratic regulation of markets and technologies. They believe in the inherently liberating power of technological innovation enabled by radical free market dynamics, and they embrace techno-utopian visions in which ensuring the bright future of humanity justifies certain sacrifices in the present. A key force behind the promotion of AGI has been the venture capital industry, with OpenAI CEO Sam Altman being a prominent example, as he is a venture capitalist himself.

Considering this context, this paper inquires about AGI as a paradigmatic example of deep hype driving a cyberlibertarian programme. Deep hype is conceptualised for the first time here as a distinct form of technological overpromising that differs from conventional hype in a crucial way: it is projected into the long-term future by making grandiose promises of civilisational transformation. This orientation into a vague future vision mobilises a multi-dimensional network of uncertainties, making its promises difficult, if not impossible, to verify in the present. Consequently, deep hype expands the generally short window of opportunity of traditional hype, sustaining excitement and ambiguity long enough to attract continuous investment and attention.

This paper argues that the uncertainties driving deep hype are articulated through sociotechnical fictions (Belsunces, 2025), a specific form distinct from cinematographic and literary fiction, which emerges within science and technology, by its legitimacy and consequently not recognised as such. Sociotechnical fictions are easier to produce than facts and often mobilised as if they were so. Hence, they make emerging entities intelligible, generating surprise and sense of wonder. Hence, they perform in orienting behaviour and decision-making, helping not-yet-existent technologies come into being.

From this perspective, AGI is considered the largest contemporary sociotechnical fiction due to its performative capacities in investment, media, and technological research. To explore how AGI deep hype relates to sociotechnical fiction, this paper conducts an analysis of a series of statements and documents from key AGI actors. These arenas are organised into two groups: constitutive uncertainties

–conceptual, application, temporal, value–, which are inherent to this emergent and unstable technology; and consequential uncertainties –governance, economic, geopolitical, existential and ethical–, which summarise the potential effects this technology might have across different societal levels.

The analysis concludes that AGI deep hype is the result of venture capital logic that converges technological and financial speculation in its functioning. Operativised through a myriad of uncertainties, the sociotechnical fiction of AGI establishes a recursive relation to deep hype: its hype is sustained by a number of uncertainties that are tamed and excited by sociotechnical fictions, while the hype and the uncertainty are fed by such fictions.

This illustrates how AGI deep hype is instrumental for mobilising resources, framing visions of the future as unavoidable, legitimises acceleration, nudges regulatory action and consolidates the cyberlibertarian and longtermist agenda. By understanding how deep hype, uncertainty, sociotechnical fictions, venture capital speculation and technopolitics converge, this theoretical paper contributes to the emerging field of hype studies[1] (Powers 2012, van Lente et al 2013, Alvial and Konrad 2016, Dedehayir and Steinert 2016; Bareis, Roßmann and Bordignon 2023, Bareis 2024) and feeds into the debate of governance of technological futures (Guston, 2008; Jasanoff and Kim, 2015; Beckert, 2016).

[1] The first Hype Studies Conference took place an the Open University of Catalunya (Barcelona, Spain) in September 2025

# Hype, uncertainty and sociotechnical fictions in techno-financial capitalism

## *Technology hype*

Technology hype has been analysed through various lenses, including consultancy research, marketing, communication studies, Science and Technology Studies (STS), and the sociology of innovation. The term gained popularity through Gartner's "hype cycle", which tracks emerging technologies through phases of inflated expectations, disillusionment, and stabilisation. Historically, this model has served as a mechanism for understanding technological change, although it has been considered as a folk theory (Alvial Palavicino & Konrad, 2016) and a scientific communication failure (Interman, 2022). It has also been criticised as inaccurate and simplistic (Dedehayir and Steinert, 2016), failing to capture the complex, contingent factors driving technological innovation while misleading decision-making.

Acknowledging its toxic potential, this paper explores a performative and political approach to hype (van Lente, Spitters and Peine, 2013), while also considering it an agent mobilised in power struggles. It treats it as an inherent force of technological emergence that is deliberately and strategically mobilised to secure economic, cultural, and political objectives. From this perspective, hype is a future-oriented, overpromissory sociotechnical phenomenon. It consists of hyperbolic discourses designed to create confidence and persuade stakeholders (Stilgoe, 2020) of the desirability and inevitability of technological change. An archetype of technological and financial capitalism, as well as western entrepreneurial culture (Wadhwani and Lubinski, 2025), hype is a manufactured event (Klingebiel, 2021) that attracts attention and generates the illusion of a rapidly closing window of opportunity (Bareis, Roßmann & Bordignon, 2023) that fosters fear of missing out (Markelius et al, 2024). It fuels speculative investment and compresses temporal horizons, pressuring actors to act before opportunities vanish. These accelerated temporalities intensify urgency, driving rapid financial commitments and reinforcing an economic environment dominated by short-term speculation.

Structurally, hype follows a trajectory of rising expectations and attention, often followed by disappointment. Short-lived and open-ended, hype sometimes leads to breakthroughs but often does not. It crystallises and instrumentalises collective visions of the future, thriving on uncertainty and producing the illusion of certainty about what is yet to come. As promotional discourse (Millar, Batalo and Budgell, 2022), hype heralds seemingly unique novel developments unfolding in a projected sequence of events.

Hype inherently engages power dynamics since it always benefits some actors while harming others. By amplifying positive implications and downplaying limitations, it legitimises technological ventures (Garud et al, 2023) while overseeing others. Often uncritically received by stakeholders, hype shapes innovation trajectories, drives investment, and structures collective behaviour across entrepreneurs, researchers, investors, policymakers, consultants, journalists, and influencers.

Often actors engage in hype because it **c**reates opportunistic dynamics (Bareis, 2024), as they seek to capitalise on fleeting moments of heightened expectations. As a persuasive strategy, hype's seductive and misleading agency operates differently across stakeholders, reinforcing existing power asymmetries in technological innovation. This happens because the ability of creating and spreading hype is unevenly distributed. Frequently championed by charismatic leaders, hype fosters collective belief. It builds communities of practice aligned around common visions and fuels competitive dynamics (van Lente, Spitters and Peine, 2013). Proliferating through sensationalist narratives (Goldfarb and Kirsch, 2019) and buzzwords (Bensaude Vincent, 2014), hype amplifies expectations and momentum. At the affective level, it generates contagious anticipation operating as a driver for desire traversed by excitement and hope at its peak, followed by fear, pessimism, frustration, and disillusionment (Wachawami and Lubinski, 2025) when it fades.

### Uncertainty & Ambiguity

Technology hype is a crucial mechanism in the sociotechnical production of futures. In western societies, the future is an endless space for uncertainty. Since the enlightenment, science and technology have been striving to face, understand and clear the uncertainties posed by the future. Paradoxically, today uncertainty is no longer simply an obstacle to overcome but has become a productive force within technoscience. Scientific and technological activities do not eliminate uncertainty; they often generate new kinds of it. Uncertainty is co-produced with temporal certainty and knowledge, making technosciences thrive in conditions of incomplete control and unpredictability. As a consequence, as technology and science increasingly interact with complex systems, novelty and unpredictability proliferate (Nowotny, 2015).

Hype seems to tame uncertainty since it creates exaggerated promises of technological results, urges actors to act, and creates pathways of normative desire. But given its volatile nature, it also stimulates it. In this regard, ambiguity performs a key role in hype in a context where financial interest and technological innovation are inherently intertwined –hence the formulation of *techno-financial*. In an emerging technology, when the mechanisms for delivering value remain undefined, sociotechnical fictions proliferate, increasing the likelihood that different investors will commit to a promissory vision, particularly when projecting to 'disrupt' markets or replace entire value chains (Goldfarb and Kirsch, 2019).

However, hype operates unevenly across technological domains. Technologies with generic or vague applications produce more diverse and productive expectations (van Lente, Spitters, and Peine, 2013). Where few facts ground belief, investors, technologists and researchers spin optimistic tales to justify decisions (Goldfarb and Kirsch, 2018, 49). Indeed, broad promise thresholds engage heterogeneous researchers, technologists, investors and citizen interest. This, in general, makes hype more resilient, allowing pivots, rebrands, or reconfigured expectations even after disillusionment. In contrast, narrowly defined technologies, like for example insulin, face deterministic hype trajectories, making failure harder to recover from.

Ambiguity plays a key role in how technology hype engages uncertainty and power. Pfaffenberger (1992) explains how ambiguity operates in the establishment of standards and rules of use, while suppressing alternative configurations of understandings of a particular technology.

Interpretative flexibility, inconsistency or even outright contractions are instrumental in achieving different political, economic, and cultural ends, readdressing sociotechnical systems such as regulations, innovation avenues, investment patterns and market dynamics.

### *The agency of sociotechnical fictions in technology hype*

Technology hype exists in an epistemic space traversed by uncertainty and suspended between truth and falsity, bridging scientific research, entrepreneurial imagination, financial speculation, and material reality. As a form of giving shape to emerging technologies, hype deeply engages with sociotechnical fictions (AUTORb). Such fictions are mediated forms of imagination that are collectively agreed upon and are part of the necessary uncertainty involved in emerging, future-oriented technological projects.

Unlike literary or cinematographic fiction, sociotechnical fictions exist within the contours of science and technology. However, they are not recognised as such because they are sheltered under technoscientific legitimacy. By making imaginary entities intelligible through metaphors, scenarios, prototypes, and technical documentation, they play a crucial role in rational and instrumental processes that attempt to bring not-yet existent technologies into being.

The notion of sociotechnical fiction expands on Beckert's fictional expectations (2016) by acknowledging its performative capacities in epistemic, aesthetic, affective and normative domains. A defining epistemic feature of such fictions is that they are more easily produced and mobilised than facts. Like hype, they flourish in uncertainty, operating as promissory 'as-ifs' that lend plausibility to emerging technologies. By structuring ambiguous futures into tangible visions, these fictions increase hype's traction among stakeholders.

Alongside their epistemic role, sociotechnical fictions perform aesthetic agencies. Since they go beyond the conventional logic chains (Latour, 2013), they enable imagination to travel far from objective knowledge and therefore foster new articulations, challenging what is taken for granted, and therefore stimulating novelty, sparking curiosity and surprise. This connects with the affective capacities of sociotechnical fictions. They bring the unknown into the understandable, exciting affects such as wonder, desire, and uncanniness. This dynamic participates in the re-enchantment (Knorr-Cetina, 1994) of actors' experience, connecting mythological and esoteric realms to empirical reality.

Sociotechnical fictions are also normative, scripting collective action by organising and orienting expectations. They establish shared frameworks that align stakeholders around common visions, reinforcing belief communities essential for hype to gain momentum. Embedded in technological discourse through metaphors like 'cloud computing', 'quantum technologies' or the 'metaverse' they become self-reinforcing, guiding investment and development trajectories (van Lente, 1993).

Thus, sociotechnical fictions are not merely auxiliary to hype, they are integral to its functioning. They sustain hype's capacity to make imagined futures understandable, and therefore engage different kinds of media, technological and financial actors to extend its reach and gain temporal legitimacy. Unlike the linear path suggested by the hype cycle, both hype and sociotechnical fictions remain potentially reactivatable, ensuring their continued influence in shaping technological futures.

### *Venture Capital's Quadruple Speculation*

Hype is neither homogeneous nor absolute. It manifests in varying degrees depending on the socioeconomic and political context, shaped by who drives it, how new entrants engage with it, and how knowledge asymmetries affect its adoption (van Lente, 1993). These dynamics intensify within contemporary techno-financial capitalism, where financial markets and organisations, as well as technology companies, play an outsized role in shaping sociotechnical trajectories. As of 2024, the financial services sector constitutes approximately 31% of the global economy[2], growing at an annual rate of 7.7%[3]. The digital economy (including AI, cloud computing, and e-commerce) accounts for an estimated 15–20% of global GDP, with 60–70% concentrated in the U.S. tech market[4]. According to these metrics, financial and technological sectors sum up almost 50% of the world economy, showing how central techno-financial capitalism is.

The connection between speculation and finances has been related, at least since the sub-prime crisis of 2008 (Pemmaraji, 2015), with the irrational, irresponsibly and

---

[2]

https://statisticstimes.com/economy/projected-world-gdp-ranking.php
[3]

https://www.researchandmarkets.com/reports/5939673/financial-services-global-market-report
[4]

https://www.bea.gov/system/files/2022-11/new-and-revised-statistics-of-the-us-digital-economy-2005-2021.pdf

predatory economic operations that exploit uncertainty to accumulate capital in the absence of any 'reliable' evidence (Ericson & Doyle, 2004). It also defines a form of exploring the unknown (Wilkie, Savransky & Rosengarten, 2017), and also an hegemonic form of social imagination (Komporozos-Athanassiou, 2022), widely mobilised in front of the uncertainty provoked by the broken promises of neoliberalism.

Speculation also plays a key role in the technology industry since it is highly future-oriented and deeply entrenched with economic speculative operations. In this convergence between financial and technological speculation lies a key actor. Venture capital (VC) is a high-risk, high-reward industry funding emerging technologies and startups with the potential for exponential growth, playing a pivotal role in shaping technological futures and consolidating economic power (Ferrary and Granovetter, 2009). Central to consolidating the hegemony of the US tech sector, VC has doubled in value since 2010 and expanded at a rate of 15–20% annually[5].

VC is speculative in at least four ways: imaginatively, financially, technologically, and ethically. VC speculative imagination is forward-thinking since it invests in risky possibilities that require the ability to anticipate which emerging technologies might shape new industries. This involves projecting market conditions far into the future and constructing narratives that align technological advancement with economic opportunity, while reproducing entrepreneurial optimism, mobilising technological foresight, and fostering a culture seeking to produce revolutionary futures epitomised by the idea of 'disruption' (Wadhwani and Lubinski, 2025).

Financially, unlike traditional models focused on stable revenue, VC thrives on funding companies with no immediate profitability, high volatility and radical growth potential —a traditional feature of speculative finances. At the technological level, VC is speculative because it bets on not-yet-existing artefacts that promise exponential returns, leveraging volatility and uncertainty as advantages. Finally, VCs are ethically speculative. By backing projects that claim to safeguard humanity's future such as cybersecurity, AI safety, biosecurity, gene therapy or space colonisation, VCs position themselves as stewards of long-term survival.

### *The politics of the venture capital speculation*

VC are central actors not only in the economic dominance of techno-financial capitalism, but also in its cultural hegemony. Entrepreneurial culture and the so-called Californian Ideology (Barbrook & Cameron, 1996) have become pervasive, positioning technological disruption as inevitable and desirable.

Since the last few years, some venture capitalists have raised their political profile. Marc Andreessen, founding partner of Andreessen Horowitz (A16), one of the wealthiest VC firms, published in 2023 the "Techno-Optimist Manifesto". This text, written by a loud supporter of US President Donald Trump, exemplifies a political agenda often labelled cyberlibertarianism (Winner 1997, Golumbia, 2024): a form of free-market fundamentalism that believes digital technologies inherently lead to freedom and progress; views democratic governments as obstacles to innovation and personal freedom, seeks to render democratic governance and regulatory frameworks obsolete; and advances visions of the future that are often exaggeratedly and blindly utopian.

---

[5]
https://nvca.org/wp-content/uploads/2023/03/NVCA-2023-Yearbook_FINALFINAL.pdf

Cyberlibertarianism, diverse in nature, encompasses overlapping positions in the US technology industry. One increasingly influential strand is longtermism, an ethical and philosophical perspective that prioritises the far future, arguing that ensuring the survival and flourishing of future generations should be a central moral concern (MacAskill, 2022), even if it causes suffering today (Torres, 2022). This thinking generally portrays concentrated wealth and power as essential to 'steering' civilisation toward optimal futures. As it will be further developed, embedding existential risk reduction into their speculations, VCs and technologists frame extreme risk-taking as an ethical duty rather than a gamble, legitimising their role as visionary architects of the future rather than mere financiers. This facilitates VC to advance with a programme that concentrates wealth and power, increases inequality, advances towards the erosion of democratic values and institutions, strengthens the economization of social relations, and deepens the already rampant assetization of everyday life (Castro & Belsunces, 2024).

### *Uncertainty, sociotechnical fictions and hype in venture capital speculation: governing technological futures by other means*

Anticipatory governance is defined as 'a broad-based capacity extended through society that can act on a variety of inputs to manage emerging knowledge-based technologies while such management is still possible' (Guston, 2008: vi). Finance and the technology industry play a key role in the governance of futures (Beckert, 2016; Birch, 2016; Birch and Muniesa, 2020).

VCs, as fundamental techno-financial actors, ground their growth model in strategically managing uncertainty through their speculative capacities. In this process, emerging technologies are made intelligible through sociotechnical fictions that align visions, spark surprise and desire, and drive investment alongside promises of substantial revenue. Sociotechnical fictions are disseminated through portfolios, white papers, and media outlets, presented as if they were already operative. This enables the suspension of disbelief and fuels the hype dynamics that VCs rely on to fund their technologies. VC speculation both boosts and is boosted by hype and sociotechnical fictions. This explains why VC investors explicitly encourage entrepreneurs to overpromise, fostering expectations that attract funding and market momentum—the so-called "fake it until you make it" myth.

As will be shown in the following sections, VC speculation plays a key role in governing the technological future by leveraging highly performative sociotechnical fictions. These thrive on uncertainty and circulate through hype, generating cycles of investment, framing what is presented as unavoidable, influencing regulatory and research agendas, shaping sociotechnical trajectories while accumulating profit and power and advancing a cyberlibertarian programme.

# Artificial General Intelligence

In January 2023 Microsoft announced a $10 billion partnership with OpenAI. This was the largest strategic alliance to date in the AI industry, and was aimed, among other objectives, to provide the company with funding and infrastructure to achieve Artificial General Intelligence (AGI). More recently, in April 2025 OpenAI raised up to $40 billion in a mega funding round led by SoftBank alongside co-investors Microsoft, Coatue, Altimeter, and Thrive at a valuation of $300 billion (Hu et al, 2025), with the explicit objective to fuel AGI research. This sets an unprecedented record since it is the largest single funding round ever raised by a private tech startup.

Paradoxically, although AGI has become a north-star goal for AI research (Blili-Hamelin et al. 2025), this term remains controversial and contested (Summerfield, 2023; Morris et al 2023; Blili-Hamelin et al, 2024; Gebru and Torres, 2024; Mitchell, 2024; Ahmed et al, 2024). Some researchers argue that current AI methods (particularly large language models) are ill-suited or incapable of achieving AGI (Landgrebe and Smith, 2019; Fjelland, 2020; Sublime, 2023; Jaeger, 2023; Schlereth, 2025). However, others claim that LLMs already embody AGI (Agüera y Arcas and Norvig, 2023).

AGI is often conflated with adjacent terms like "human-level AI," "superintelligence" (Bostrom 2014), and "strong AI" (Smart, 2015), broadly understood as systems capable of automating and overcoming (Chalmers, 2010) human cognitive functions, learning autonomously (Chollet et al. 2024), manipulating general knowledge (Goertzel 2014), solving novel problems, and economically outperforming humans (Suleyman and Bhaskar, 2023; OpenAI, 2018). Some definitions extend further, pointing even to consciousness (Smart, 2015).

Speculative philosophers aligned with Silicon Valley corporations have framed the rise AGI in in quasi-divine terms, stating that "nonbiological intelligence created in that year [2045] will be one billion times more powerful than all human intelligence today" (Kurzweil 2005, p. 136) or that it would "likely exceed humans in every cognitive domain" (Bostrom 2014, p.22). In this regard, technologist and investor Ian Hogarth stated that "God-like AI" could lead to the "obsolescence or destruction of the human race" (Syme, 2023).

Logically, AGI has been described as "purely a marketing term, disconnected from the research sphere" (AI Now 2024, 23), strategically deployed to distinguish new products from the broader field of AI research. Functioning as a normative technological metaphor (Wyatt, 2016), AGI shapes expectations and structures discourse well beyond its technical feasibility. As a buzzword, AGI fits the characteristics outlined by Bensaude-Vincent (2014): hyperbolic, rapidly circulating, generating matters of concern, building consensus, and setting expansive but often vague goals.

### The politics of AGI

The notion of AGI was coined by Ben Goertzel and Cassio Pennachin (2007). Goertzel is a vocal cyberlibertarian and an early member of the transhumanist movement. He is critical of both the centralised democratic state and corporate monopolies, advocating instead for crypto-anarchist and decentralised forms of governance enabled by blockchain technologies. This longtermist orientation is shared by other prominent AGI proponents, such as philosopher Nick Bostrom and singularity advocates like Ray Kurzweil and Eliezer Yudkowsky. However, unlike Goertzel, they tend to adopt more alarmist views and support centralised approaches to AGI safety.

The idea of AGI did not become mainstream until 2018 when OpenAI legitimised the term within corporate, investor, and media circles by asserting that its mission was to ensure that AGI benefits "all of humanity" (OpenAI, 2018). Sam Altman, OpenAI's co-founder and CEO, is first and foremost an investor who has spent more than half of his professional career in the venture capital industry. Other initial investors of this company have been cyberlibertarian tech leaders such as Elon Musk and venture capitalist and loud neo-reactionary Peter Thiel (Gebru and Torres, 2024). As this paper shows, this background helps explain what is the political programme reproduced by AGI, and how sociotechnical fictions and hype contribute to it.

Just like the popular conception of AI, AGI is grounded in science fiction imaginaries and draws on both contemporary and ancient myths. At the technoscientific level, it is marked by controversy and a lack of precise definition. In this light, AGI functions as a large-scale sociotechnical fiction: an imagined entity presented and supported by science and technology *as if* it were an inevitable fact, while remaining speculative. Crucially, sociotechnical fiction is far easier to mobilise than facts, which poses an instrumental quality for actors invested in sustaining hype. By borrowing credibility from generative AI's visible progress, the sociotechnical fiction of AGI gains material traction, aligning stakeholders around a vision that frames its development and control as a societal necessity, and that is powered through uncertainty and hype dynamics, driven and profited from by venture capital speculation to gain momentum. It is precisely this combination that positions AGI as a specific case of AI hype.

Recent critiques of AI hype highlight its growing political, economic, and social consequences. The U.N. Advisory Board warns that "an overwhelming amount of information (...) makes it difficult to decipher hype from reality", fuelling confusion and reinforcing the power of major AI companies over policymakers and the public (United Nations, 2024). Floridi (2024) situates AI hype within the recurring logic of speculative tech bubbles, from the Dot-Com boom to cryptocurrency, where financial speculation outpaces technological reality. Along these lines, Markelius et al. (2024) argue that AI hype is historically unparalleled, generating significant planetary and social costs while obscuring the material, environmental, and labour conditions underpinning AI systems. Thais (2024) further contends that exaggerated claims distort research priorities, mislead public understanding, and lead to the premature deployment of unreliable technologies.

Acknowledging these critiques, this paper widens the scope of AI hype research, by proposing the concept of deep hype to capture how AGI sustains speculative momentum by mobilising sociotechnical fictions and uncertainty, while understanding such processes as a form of governance of AI futures.

## The Uncertainties of Artificial General Intelligence

In recent history, corporations and think tanks relied on scientific legitimacy to manufacture uncertainty around tobacco, climate change and other issues to orient public action or delay regulation (Oreskes and Conway, 2010). This uncertainty is helpful to keep investment and research avenues open, and as this paper shows, can be mobilised along with hype to consolidate technopolitical and economic power.

Hype dynamics typically emerge around specific technological breakthroughs or products, generating intense but short-lived excitement. Given its nature, AGI hype operates differently. Before analysing its uncertainties, a tentative definition of deep hype can be proposed. Deep hype is a distinct form of technological overpromising that differs from conventional hype in a crucial way: it is a long-term, overpromissory dynamic that constructs visions of civilisational transformation through a network of uncertainties extending into an undefined future, making its promises nearly impossible to verify in the present. Unlike traditional hype, which operates within shorter cycles, deep hype expands the window of opportunity, sustaining excitement and ambiguity long enough to attract continuous investment and attention in an effort to realise its grand claims. Because deep hype thrives on uncertainty, it mobilises sociotechnical fictions to make the unknown intelligible. These fictions produce the illusion that the not-yet-existent is already within reach, sparking surprise, excitement and wonder, and articulating desires or fears that push actors to invest, research, regulate, or simply pay attention.

Based on the analysis of a series of statements and documents from key AGI actors, a list of uncertainty arenas will be constructed to understand AGI deep hype. These arenas are organised into two groups: constitutive uncertainties –conceptual, application, temporal, value–, which are inherent to this emergent and unstable technology; and consequential uncertainties –governance, economic, geopolitical, existential and ethical–, which summarise the potential effects this technology might have across different societal levels.

### Conceptual uncertainty

As already mentioned, the notion of Artificial General Intelligence is subject to persistent controversy. Therefore, its first constitutive uncertainty is conceptual. Importantly, most AGI papers do not define what "general purpose" actually means (Paolo et al, 2024), and whenever is defined, there are significant differences over "how much 'generality' is desirable" (Blili-Hamelin et al, 2024). Logically, defining what 'general' means, when that is the word that upcycles one of the most famous acronyms of the XXIst century, is a necessary precondition to orient research towards this goal. However, and given the ambiguity of the concept, this lack of agreement is likely to persist in the coming years (Blili-Hamelin et al, 2025; Summerfield, 2023). Although this risks undermining scientific research quality and legitimacy, it's benefitting investors and corporations by keeping an open promissory framework for the AGI hype —which makes it deep in its reach.

### Application uncertainty

The interpretative flexibility of AGI brings its second constitutive uncertainty: what its applications will be. Unlike narrow AI, designed for specific tasks, AGI is imagined as capable of doing almost anything, yet what those tasks might be, or how AGI would integrate into existing sociotechnical systems, remains vague. This allows AGI to be framed both as the ultimate solution to global challenges and as a potential existential threat.

Proponents claim that AGI will revolutionise the economy or science. For example, OpenAI's mission is to ensure that "artificial general intelligence (...) benefits all of humanity", claiming that AGI "could help us elevate humanity by increasing abundance, turbocharging the global economy, and aiding in the discovery of new scientific knowledge" (OpenAI, 2023), or more recently that "scientific progress will likely be much faster than it is today; this impact of AGI may surpass everything else" (2025b). Some even frame AGI as the key to solving challenges like climate change (Smitsman et al, 2024). Similarly, Demis Hassabis, CEO of DeepMind asserts that AGI "can solve what I call root-node problems in the world—curing terrible diseases, much healthier and longer lifespans, and finding new energy sources. If that all happens, we [will] travel to the stars and colonize the galaxy" (Levy, 2025).

AGI's vague promises are sustained by sociotechnical fictions and aligned with longtermist visions. However, the expectations they generate are also animated by "proofs of verifiable truth" (Jasanoff & Kim, 2015) that strengthen belief in specific future visions. For instance, OpenAI's scaling laws paper (Kaplan et al, 2020) suggested that increasing model size improves performance—a claim reinforced by ChatGPT's success. This, along with the generative AI boom and its rapid adoption blurred the line between fiction and fact, fuelling hype and legitimising AGI visions.

In this sense, the openness to the possible applications of AGI contributes to the depth of its hype.

*Temporal uncertainty*

Another key constitutive uncertainty in AGI is its temporal threshold: when this technology will emerge. Predictions about AGI timelines vary significantly, reinforcing uncertainty. In 2011, Shane Legg, DeepMind's co-founder estimated a 50% chance of AGI by 2028 (Legg, 2011). In September 2024, Altman spoke of "a few thousand days" to reach superintelligence, while a month later, Anthropic CEO Dario Amodei (2024) claimed it could arrive as early as 2026—or much later. By January 2025, Altman asserted OpenAI knew how to build AGI and predicted AI agents would soon integrate into the workforce (Altman, 2025a). In June 2025, Hassabis asserted that AGI "will begin to happen in 2030" (Levy, 2025).

Meanwhile, some researchers argue that AGI "sparks" are already visible (Bubeck et al, 2023), while others insist that current AI methods are structurally incapable of achieving general intelligence (Landgrebe and Smith, 2019; Fjelland, 2020; Sublime, 2023; Jaeger, 2023; Schlereth, 2025). These shifting timelines often reflect funding cycles, competitive pressures, and the need to maintain leadership in the AI race.

These temporal contradictions create a paradox where AGI is near and long-term at the same time. On the one hand, AGI is framed as imminent, requiring swift investment and regulatory laissez faire. On the other, it remains uncertain and distant, positioned as a transformation that could take decades or more. This dual framing sustains AGI as a long-term inevitability while justifying present-day decisions aligned with the interests of those driving the hype. This allows promoters to avoid short-term accountability while keeping investment and attention fixed on an ever-receding technological horizon. This influences the temporal dynamics that nudge the governance of AI futures.

*Value uncertainty*

An added core uncertainty shaping AGI deep hype is its economic value. Unlike other AI technologies that generate revenue through specific applications, AGI lacks a well-defined business model or clear path to profitability (AI Now, 2024). OpenAI's origins as a nonprofit reflected this uncertainty, as revenue generation was not initially a priority. This changed with its transition to a capped-profit model, underscoring the broader ambiguity surrounding AGI's financial sustainability.

Reports indicate that OpenAI and Microsoft have set a financial benchmark for AGI, defining its achievement not through technical milestones but by generating at least $100 billion in profits (Wiggers 2024). This reveals how, at least in part, AGI's economic rationale rests less on proven capabilities and more on long-term financial speculation. In this sense, Altman alternates between downplaying AGI's impacts when asserting that "AGI can get built, the world mostly goes on in mostly the same way, things grow faster" (Pillay, 2025), while at the same time assuring investors that returns will materialise once the technology is fully developed.

This ambiguity is no accident. AGI has been shaped by venture capital speculative dynamics. It enables exaggerated claims that AGI will revolutionise industries and unlock exponential growth despite no evidence that such transformation is achievable.

Yet, its development demands massive speculative investment with uncertain returns, a risk venture capital embraces to secure privileged positions in AI's future. The way AGI leaders are framing the agenda and orienting of economic and media actors operates as a clear strategy of governance of AI futures.¡

<u>Consequential</u>

Having examined the constitutive uncertainties that sustain AGI's deep hype: its conceptual ambiguity, application vagueness, shifting timelines, and speculative value, what follows focuses on the projected consequences of AGI's realisation. These consequential arenas are shaped by hopes and fears, structured by sociotechnical fictions that envision AGI as both a civilisational breakthrough and a potential catastrophe. Grounded in the ambiguity encompassing any future scenario, the depth of AGI hype performs at governance, economic, geopolitical, existential and ethical levels.

*Governance uncertainty*

Since AGI is presented as a force capable of civilisational transformation, it consequently generates profound uncertainties regarding power, governance, and the (un)democratic shaping of technological futures. For example, on 16 May 2023, Altman testified before the U.S. Senate during a hearing titled "Oversight of A.I.: Rules for Artificial Intelligence". The same hearing also included testimonies from Christina Montgomery, Chief Privacy and Trust Officer at IBM, and Gary Marcus, Professor Emeritus at New York University (U.S. Senate 2023). In this context, each participant advocated a different regulatory approach—some more stringent than others. Altman warned about the dangers posed by AGI, while Marcus argued that such consequences remain distant, stating that "we're not that close to artificial general intelligence, despite all of the media hype" (Hendrix, 2023). Six days after the hearing, on March 22 of the same year, the Future of Life Institute, an institution that has been generously funded by Elon Musk, published an open letter entitled "Pause Giant AI Experiments" calling to pause on training AI systems more powerful than GPT-4 due to the profound risks to society and humanity. These events unleashed a wave of worldwide media coverage (Bareis, 2024) warning about the possible dangerous consequences of AI. An example of that is Yudkowsky's (2023) article in Time magazine entitled "Pausing AI Developments Isn't Enough. We Need to Shut it All Down".

On the other side of the regulatory spectrum, venture capitalists and free-market fundamentalists such as Marc Andreessen and Peter Thiel, along with Ben Goertzel and Ethereum co-founder Vitalik Buterin, hold anti-regulation or regulation-sceptical positions regarding AI and technology.

These brief examples show how the controversial debate around AI and AGI regulation entangles uncertainty, hype and the governance of AI futures. By leveraging uncertainty, testifying before governments, proposing AI safety frameworks, writing manifestos or participating in open letters, venture capitalists and tech elites exert their influence by shaping or eroding regulation while maintaining their economic power.

The uncertainty surrounding AI futures is not only a matter of anticipatory governance but is also strategically leveraged as a business opportunity. Worldcoin, another of Altman's ventures, presents itself as an identity and financial infrastructure for the AI future, claiming to "empower everyone to participate in the global economy in the age of AI and provide the foundation for shared governance" (Worldcoin 2023). In this case, the potential dangers posed by AGI technologies developed by Altman are purportedly addressed by another technology also created by him, enclosing both the problem and the solution within the same entrepreneurial logic. This centralised and corporate-led position on safety-focused AI governance has been criticised by Goertzel (Fagella, 2023), the person who coined the term AGI and who's openly against AGI regulation.

Although diverse in their positions, these actors situate themselves not just as industry leaders but as architects of the AI future, embedding their cyberlibertarian and longtermist positions into AGI's imagined future, promoting technocracy, and oligopolistic control over democratic accountability. The recent authoritarian turn among parts of the US tech elite reflects this drift (Späth, 2025).

*Economic uncertainty*

Another uncertainty unleashed by AGI is related to its economic consequences, particularly around productivity, labour, and the restructuring of economic systems. AGI is imagined as a transformative force that could reduce production costs and overcome material limits altogether. Altman asserts that AGI will unlock post-scarcity abundance, claiming that "the price of many goods will eventually fall dramatically" (2025b) as AGI integrates into economic systems.

Yet, these promises raise critical questions about how AI might prioritise capital over labour and amplify the already existing inequalities (Lowitzsch and Magalhães, 2025). While some argue it might complement human work if well regulated (Pasquale, 2022), others warn it could accelerate automation and displace jobs. Paradoxically, and despite the overoptimistic statements by Altman, OpenAI published a working paper in collaboration with the University of Pennsylvania (Eloundou et al, 2023) describing the important labour implications of ChatGPT, contributing to the already rampant "hysteria" created by the rise of generative AI products (Grady and Castro, 2023).

At the core of these economic projections lies the belief that AGI will inevitably deliver vast gains, while diminishing the costs of labour. Here, the combination of hype, the production of uncertainty and its management through sociotechnical fictions —for example, stating that AGI will bring unprecedented abundance—, presents its future as unavoidable while framing any resistance as futile. This exacerbates hopeful and fearful reactions that affect the political debate. In doing so, it preempts alternative economic models and solidifies the expectation that AGI-driven productivity will reshape economies, regardless of its feasibility or social costs.

### Geopolitical uncertainty

AGI is also framed as a geopolitical and security concern. Military and policy circles, like for example the military think tank Rand Corporation, compare AGI to nuclear technology, warning of an "AGI arms race" between the US and China (Mitre and Predd, 2025; Marshall, 2025). Despite the doubts about whether and when AGI will become feasible, these narratives cast it as a potential tool for global dominance, driving competition and positioning AGI developers as strategic assets.

This scenario justifies aggressive investment and enables regulatory capture, where private actors shape policies to entrench their control. Grand geopolitical projects like *Stargate*—a $500 billion initiative named after a science fiction franchise backed by tech giants and endorsed by President Trump—show how AGI hype and its related sociotechnical fictions materialises in large-scale industrial infrastructures and energy-intensive projects, justified by the need to stay competitive in the AI race.

### Existential uncertainty

As previously explained, AGI has been conceptualised as a longtermist vehicle to help humanity overcome grand challenges and advance toward a new civilisational stage. Goertzel, in its singularitarian and transhumanist *Cosmist Manifesto* (2010, pp. 10–11), envisions that AGI will enable mind uploading, allowing "indefinite lifespan to those who choose to leave biology behind and upload." Humans, he claims, "will spread to the stars and roam the universe," create an "abundance of wealth and growth," and ultimately reach "states of individual and shared awareness possessing depth, breadth and wonder."

Opposed to these techno-utopian visions, other cyberlibertarians from the industry and the academia warn about AGI as an existential risk: prominent figures like speculative philosopher Nick Bostrom (2014), technologist and writer Ray Kurweil, and AI researchers Eliezer Yudkowsky (Dexa, 2023) and Geoffrey Hinton (Webb, 2025) warn that AGI poses a potential (or already unsolvable) threats to humanity's survival since it could surpass human intelligence and spiral out of control, leading to a potential disaster. Since these statements are often presented under technoscientific legitimacy, they perform as sociotechnical fictions.

In front of the uncertainty leveraged by AGI, such fictions operate as necessary safeguards, but since they are mobilised by an ideology that is in general contrary to democratic institutions, they also function to justify the concentration of control over AGI in the hands of a few. This raises important questions about who benefits from circulating these fears and how they shape decisions at societal, economic, and governance levels.

In this respect, existential risk and its inherent sociotechnical fictions operate as "wishful worries" (Brock, 2019) where imagined threats overshadow current and urgent problems. By projecting extreme futures, AGI hype redirects attention and resources, consolidates control, and reinforces calls for unregulated progress. The dual narrative of humanity's salvation versus an apocalyptic threat creates the illusion of AGI as urgent and unavoidable, regardless of its technical feasibility.

*Ethical uncertainty*

The uncertainties listed below provide a stage for corporations and investors involved in developing AGI, whose role extends beyond creating new technologies and stimulating markets. It is also framed as an ethical obligation in response to alleged existential risks or as a catalyst for civilisational transformation. These positions are grounded in uncertain futures and longtermist ethics, which justify present-day sacrifices in the name of a supposedly greater tomorrow.

For example, OpenAI's (2023) mission is presented in ethical terms when it states that it wants AGI to be "an amplifier of humanity" while navigating "massive risks". On this matter, this company recently stated that it aims at creating guardrails to ensure responsible AI development in its 'EU Economic Blueprint' report (2025).

From a more anti-regulatory and technocratic perspective, Andreesseen stated that society should do everything possible to accelerate technology development and advancement without constraints since "there are scores of common causes of death that can be fixed with AI, from car crashes to pandemics to wartime friendly fire". Hence, developing AI becomes a longtermist ethical obligation since "any deceleration of AI will cost lives. Deaths that were preventable by the AI that was prevented from existing [are] a form of murder" (2023).

Such overoptimistic and longtermist ethical claims –articulated through sociotechnical fictions– ignore other critiques stating that AI will deepen inequalities, reshape social hierarchies, and undermine existing knowledge structures (Eubanks, 2018; Broussard, 2018), while having current negative externalities by rising energy and water consumption, among others.

# Discussion: AGI Deep Hype, Sociotechnical Fictions and the Governance of AI Futures

*AGI Deep Hype*

Deep hype is a distinct form of technological overpromising that differs from conventional hype in a crucial way: it is a long-term, overpromissory dynamic that constructs visions of civilisational transformation through a network of uncertainties extending into an undefined future, making its promises nearly impossible to verify in the present. Unlike traditional hype, which operates within shorter cycles, deep hype expands the window of opportunity, sustaining excitement and ambiguity long enough to attract continuous investment and attention in an effort to realise its grand claims. This dynamic creates a structure of shared technological belief — not unlike faith— that shapes investment, policy, and public discourse, often without delivering tangible progress.

Because deep hype thrives on uncertainty, it mobilises sociotechnical fictions to make the unknown intelligible. These fictions produce the illusion that the not-yet-existent is already within reach, sparking surprise, excitement and wonder, and articulating desires or fears that push actors to invest, research, regulate, or simply pay attention. This combination of depth and ambiguity creates a carefully maintained balance between risk and certainty, sustaining momentum and feeding venture capital's appetite for exponential returns.

AGI is a paradigmatic example of deep hype. Like nuclear fusion, quantum computing, or brain-computer interfaces, AGI is perpetually positioned as just beyond reach and at the same time a force of civilisational change. Te. This ambiguous timeline allows proponents to assert both inevitability and unpredictability, reinforcing technopolitical assumptions, generating self-sustaining narratives that justify continued investment, and shaping the governance of technological futures.

AGI uncertainty creates an environment in which hype operates both as an investment vehicle and as a mechanism for shaping technological trajectories. The fact that AGI has attracted the largest private funding round in history, despite the absence of consensus on its definition or technical feasibility, and amidst a myriad of unresolved uncertainties, underscores its speculative nature in both financial and technological terms. This makes AGI the most prominent contemporary sociotechnical fiction. This phenomenon can be partly explained by the logic of venture capital speculation, epitomised by figures like Sam Altman, capable of mobilising massive amounts of investment and media attention based on vague societal and economic promises.

Unlike conventional hype cycles driven by technological breakthroughs and rapid returns, AGI deep hype is structurally different. AGI shares many characteristics with broader AI-driven hopes and fears but differs in some key points that explain the depth of its hype. First and foremost, while relying on the AI hype powered by tangible results, AGI remains almost entirely speculative. Also, unlike AI, which has been conceptualised and developed by scientists and researchers that were initially funded by democratic states, AGI is not rooted in empirical scientific progress but in speculative thought, financial interests and technopolitical agendas. Its current popularity cannot be understood without the lever of the venture capital speculation, shaped by cyberlibertarian ideologies projected towards the longtermist technocratic future.

AGI hype is deep in at least three senses: it unleashes promises of deep civilisational transformation; its possible effects are deep and complex; and time in which this technology will operate is deep — to the point of being presented as the key for an interplanetary humanity conquering the universe.

Likewise, the deepness of the hype is explained because it is sustained by a series of overlapping uncertainties. Its inherently vague 'generality' allows an expansive repertoire of expectations (van Lente, Spitter and Peine, 2013), supporting contradictory visions of what this not-yet-existent technology can do. This interpretative flexibility enables narrative renewal while offering great financial appeal.

Its deepness is also explained through the paradox that AGI stakeholders are creating a sense of urgency without the illusion of a rapidly closing window of opportunity by setting an open, mid or long-term temporal threshold, which maintains the interest preventing the disappointment to deflate the hype. This is possible due to the tension created between techo-utopian and apocalyptic discourses, in combination with the sensationalist narratives of tech leaders, the media frenzy and the massive amounts of investment that AGI is attracting.

In comparison to traditional approaches to hype, another key particularity of AGI deep hype is that is fueled by sensationalist narratives that thrive both through overpromising the positive impacts and exaggerating the negative ones. In this regard, engaging with the framework of existential risk articulates a list of wishful worries that legitimise tech leaders to appear as ethical guards of the AI future. This makes AGI deep hype a form of soft anticipatory governance of AI futures by influencing both US government and the public opinion towards cyberlibertarian and longtermist regulatory frameworks that do not question the economic and power concentration that implies this possible technology.

All this also exemplifies how AGI hype functions as a mechanism for consolidating and expanding the hegemony of tech leaders and venture capital, which in recent years has increasingly aligned with far right authoritarian principles. This is evident in the convergence of massive investment rounds, exaggerated statements, regulatory influence, the construction of large scale infrastructures such as the Stargate Project, and the capacity of these actors to shape media and research agendas. Combined with the longtermist framework, this dynamic displaces near term concerns such as algorithmic bias, environmental impact, and labour displacement. In this context, the uncertainty mobilised by AGI stakeholders continues a long lineage of ambiguity being weaponised to impose specific scripts and protocols, while suppressing alternative understandings of a given technology (Pfaffenberger, 1992). As such, this entanglement of technopolitics, uncertainty and sociotechnical fictions undermines the possibility of an accountable and democratic debate on AGI, therefore operating as a mechanism of governance of AI futures .

### The role of Sociotechnical Fictions in AGI deep hype

As Beckert (2016) explains, fictional expectations fill the epistemic void when economic decision making faces future oriented uncertainty. Focusing on AGI, this paper argues that a different yet complementary kind of fiction is mobilised in the face of long term and deeply entangled uncertainties. Sociotechnical fictions are not recognised as such, as they are shielded by technoscientific legitimacy. They are presented as if they were factual knowledge and, as a result, exert performative agency by contributing to the materialisation of not yet existing technologies. As stated earlier, AGI functions as a sociotechnical fiction precisely because it is not technically feasible, and because it is sustained by speculative statements, investment, desire, fear, and greed. Operativised through a myriad of uncertainties, the sociotechnical fiction of AGI establishes a recursive relation to deep hype: this hype is sustained by a number of uncertainties that are tamed and excited by sociotechnical fictions, while the hype and the uncertainty are fed by such fictions.

Indeed, AGI is in itself composed of many other sociotechnical fictions, such as the claim that it is just around the corner (Shane, 2011; Altman, 2024; Amodei, 2024), despite the absence of any certainty about its timeline. Other longtermist fictions claim that AGI will overcome material limits altogether and "unlock post-scarcity abundance" (Altman 2024), drastically reducing the price of all goods (Altman, 2025a). Even more radical forms of sociotechnical fictions circulate, including the belief that AGI will become a god-like entity (Syme, 2023), discover new energy sources, enable human colonisation of the galaxy (Levy, 2025), or that it may ultimately lead to human extinction (Legg, 2023).

Such fictions are mobilised through tech leaders' blog posts, interviews, media articles, books, marketing campaigns, conferences, academic articles, public hearings and technical documents, among others. Through these means, perceived as spaces for objective knowledge, sociotechnical fictions perform epistemic, aesthetic, affective, and normative agencies. Epistemically, they render imaginary futures tangible and stabilise promises, shifting AGI from nothingness to 'thingness'. Operating as powerful *as if,* these fictions align imaginaries and expectations while producing epistemic toxicity, as researchers chase the undefined, overpromissory north-star of AGI (Blili-Hamelin et al, 2025) . Aesthetically, AGI's fiction bridges the non-existent and the existent, producing novelty and surprise creating public fascination. Affectively, the interplay of hopeful and fearful uncertainties sustains urgency, keeping AGI perpetually almost-here—as both a breakthrough and a threat. Normatively, these fictions manufacture the illusion of consensus around AGI's feasibility, making the speculative appear inevitable and shaping research, investment, and policy while advancing with the cyberlibertarian agenda.

This network of sociotechnical fictions re-enchants technologically-driven social imagination, propelling it into territories once reserved for science fiction. The deepness of AGI hype connects rational, instrumental practices with the esoteric. In this regard Blili-Hamelin et al. (2025) caution the interpretative flexibility of AGI facilitates its myth-making effect, which risks amplifying unscientific thinking about AI, further entrenching the illusion of its uncontrollable status.

This ambiguity, although detrimental to science, benefits markets. As the fifth century text The Art of War (Sun Tzu, 2002) suggests, today transformed into a classic in business schools, in the midst of chaos, there is also opportunity. Indeed, cultivated ambiguity allows venture capitalists to maximise speculative gains and maintain control. This dynamic fuels urgency, compelling investors, policymakers, and publics to secure positions in a future that remains undefined but presented as inevitable.

Consequently, sociotechnical fictions are instrumental to AGI deep hype since they make intelligible a set of speculative visions, while at the same time taming and stimulating uncertainties to keep exciting the imagination of technologists, investors and the media, in a delicate balance between fear and hope. Such fictions articulate the instrumentalisation of uncertainty by technology and financial actors in two ways. First, the promise of revolutionary breakthroughs, economic revenue and geopolitical dominance accelerates funding and research, even as advancements remain theoretical. Second, the potential threats—articulated through wishful-worries and existential risks—capture attention, create ethical legitimacy, and justify the hegemony of powerful players while opening speculative business opportunities.

Through the staging of a dense network of uncertainties, sociotechnical fictions, fears and hopes, AGI deep hype participates in the governance of AI futures by opaquing, framing and controlling what can be stated about what is yet to come. Deep hype narrows the political debate, presenting its future as unavoidable, conducting a regulatory capture while framing any resistance as futile or anti-progress. In doing so, it preempts alternative economic models and solidifies the expectation that AGI-driven productivity will reshape economies, regardless of its feasibility or social costs.

Being conceptualised by radical cyberlibertarians like Goetzel and popularised by tech barons like Altman, AGI is less the product of scientific progress than of financial and political agendas aimed at securing control over imagined futures. Following Pfaffenberger (1992), the way that AGI stakeholders strategically mobilise ambiguity or contradiction resonates with previous practices that used it to establish contradiction and open areas that are normatively indeterminate to redress regulations, investment avenues and power structures. This, added to the fact that private corporations —those who are defining the AGI future— are undemocratic institutions that operate without public accountability (Feenberg, 2002).

## Conclusion

This article analyses contemporary AGI, a not-yet-existing technology that is popularised by OpenAI and is publicly and contradictorily debated by prominent venture capitalists and tech leaders, as a paradigmatic case of deep hype. Deep hype is a long-term overpromissory dynamic constructing promises of civilisation transformation that is articulated by a myriad of uncertainties projected into an undefined future – making its promises nearly impossible to verify in the present. This specific kind of hype relies on a network of sociotechnical fictions to keep the window of opportunity open to sustain the necessary ambiguity, excitement and attention over time in order to mobilise enough investment and talent to make this grand technological future come into being.

This dynamic creates a structure of shared technological belief that drives investment, policy, and public discourse, often without delivering tangible progress, operating as a speculative force in the governance of technological futures.

Based on a series of statements and documents from AGI leaders, this paper analyses nine arenas of uncertainty related to AGI. These arenas are structured in constitutive and consequential uncertainties, and consider its conceptual, application, temporal, value, governance, economic, geopolitical, existential and ethical dimensions. Based on this analysis, it argues that AGI is an emblematic case of deep hype since it is perpetually positioned as just beyond reach and at the same time a force to bring humanity to a new era. Since it is a deeply contested category with serious controversies about its feasibility, this paper understands AGI as a large-scale sociotechnical fiction. Such fictions are mobilised within the contours of science and technology but are not recognised as such, and therefore are instrumental to bring not yet existent technologies into being.

Interestingly, AGI is set within a background that is reinforced by the very narratives surrounding AGI, that ranges from techno optimistic futures to apocalyptic fears. Such context is mobilised in order to create a paradoxical sense of urgency set in an ambiguous temporal framework, while at the same time intensifying its uncertainties – that are both stimulated and tamed by other sociotechnical fictions.

Read as the result of cyberlibertarian and longtermist imaginaries, as well as articulated through the venture capital speculation epitomised by Sam Altman – the founder of the company who mainstreams AGI, this paper explores how the combination of deep hype, fear, hope, ambiguity, uncertainty and sociotechnical fictions are key in exciting investment, nudging policy, orienting public discourse, and reinforcing private control. This operates as a means of governance of AI futures while advancing with an anti-democratic technopolitical programme.

## Use of AI

# Bibliography

Ahmed, S., Jaźwińska, K., Ahlawat, A., Winecoff, A., & Wang, M. (2024). Field-building and the epistemic culture of AI safety. *First Monday*. https://firstmonday.org/ojs/index.php/fm/article/view/13626

Altman, S. (2025a, June 10). *The gentle singularity*. Sam Altman's Blog. https://blog.samaltman.com/the-gentle-singularity

Altman, S. (2025b, February 9). *Three observations*. Sam Altman's Blog. https://blog.samaltman.com/three-observations

Alvial Palavicino, C., & Konrad, K. (2016). How technology consultants assess the graphene and 3D printing hype. *SocArXiv*. https://doi.org/10.31235/osf.io/uv72r

Amodei, D. (2024). *Machines of loving grace*. https://darioamodei.com/machines-of-loving-grace

Andreessen, M. (2023). The techno-optimist manifesto. *a16z*. https://a16z.com/the-techno-optimist-manifesto/

Bareis, J. (2024). Ask me anything!😈 How ChatGPT got hyped into being (No. jzde2_v1). *Center for Open Science*. https://doi.org/10.31234/osf.io/jzde2

Bareis, J., Roßmann, N., & Bordignon, F. (2023). Technology hype: Dealing with bold expectations and overpromising. *TATuP – Zeitschrift für Technikfolgenabschätzung in Theorie und Praxis, 32*(3), 10–71. https://doi.org/10.14512/tatup.32.3.10

Barbrook, R., & Cameron, A. (1996). The Californian ideology. *Science as Culture, 6*(1), 44–72. https://doi.org/10.1080/09505439609526455

Beckert, J. (2016). *Imagined futures: Fictional expectations and capitalist dynamics*. Harvard University Press.

Belsunces, A. (2025). Sociotechnical Fictions: The Performative Agencies of Fiction in Technological Development. *Science & Technology Studies*.

Bensaude Vincent, B. (2014). The politics of buzzwords at the interface of technoscience, market and society: The case of 'public engagement in science'. *Public understanding of science, 23*(3), 238-253.

Birch, K. (2017). Rethinking value in the bio-economy: Finance, assetization, and the management of value. *Science, Technology, & Human Values, 42*(3), 460–490. https://doi.org/10.1177/0162243916661633

Birch, K., & Muniesa, F. (Eds.). (2020). *Assetization: Turning things into assets in technoscientific capitalism*. MIT Press.

Blili-Hamelin, B., Hancox-Li, L., & Smart, A. (2024). Unsocial intelligence: An investigation of the assumptions of AGI discourse. In *Proceedings of the AAAI/ACM Conference on AI, Ethics, and Society* (Vol. 7, pp. 141–155). https://ojs.aaai.org/index.php/AIES/article/view/31625

Blili-Hamelin, B., Graziul, C., Hancox-Li, L., Hazan, H., El-Mhamdi, E. M., Ghosh, A., Heller, K., Metcalf, J., Murai, F., Salvaggio, E., Smart, A., Snider, T., Tighanimine, M., Ringer, T., Mitchell, M., & Dori-Hacohen, S. (2025). *Stop treating 'AGI' as the north-star goal of AI research*. arXiv. https://arxiv.org/abs/2502.03689

Bostrom, N. (2014). *Superintelligence: Paths, dangers, strategies*. Oxford University Press.

Brock, D. C. (2019, July 25). *Our censors, ourselves: Commercial content moderation*. *Los Angeles Review of Books*. https://lareviewofbooks.org/article/our-censors-ourselves-commercial-contentmoderation/

Broussard, M. (2018). *Artificial unintelligence: How computers misunderstand the world*. MIT Press.

Castro, A., & Belsunces, A. (2025). Cryptofinancial imaginaries: how neoliberal theories are materialized in the technical principles of cryptocurrencies. *Journal of Cultural Economy*. https://doi.org/10.1080/17530350.2024.2436869

Castro, D., & Grady, P. (2023, May 1). *Tech panics, generative AI, and the need for regulatory caution* [Policy report]. Center for Data Innovation. https://datainnovation.org/2023/05/tech-panics-generative-ai-and-the-need-for-regulatory-caution/

Committee on the Judiciary, Subcommittee on Privacy, Technology, and the Law. (2023, May 16). *Oversight of A.I.: Rules for artificial intelligence* [Hearing transcript]. U.S. Senate. https://www.judiciary.senate.gov/committee-activity/hearings/oversight-of-ai-rules-for-artificial-intelligence

Dedehayir, O., & Steinert, M. (2016). The hype cycle model: A review and future directions. *Technological Forecasting and Social Change, 108*, 28–41. https://doi.org/10.1016/j.techfore.2016.04.005

Eloundou, T., Manning, S., Mishkin, P., & Rock, D. (2023). *GPTs are GPTs: An early look at the labor market impact potential of large language models*. arXiv. https://arxiv.org/abs/2303.10130

Ericson, R., & Doyle, A. (2004). Catastrophe, risk, insurance and terrorism. *Economy and Society, 33*(2), 135–173. https://doi.org/10.1080/0308514042000225733

Eubanks, V. (2018). *Automating inequality: How high-tech tools profile, police, and punish the poor*. St. Martin's Press.

Faggella, D. (2023, July 4). *Ben Goertzel on the rise of decentralized AGI*. danfaggella.com.
  https://danfaggella.com/goertzel1/

Feenberg, A. (2002). *Transforming technology: A critical theory revisited*. Oxford University
  Press.

Ferrary, M., & Granovetter, M. (2009). The role of venture capital firms in Silicon Valley's
  complex innovation network. *Economy and society*, *38*(2), 326-359.

Fjelland, R. (2020). Why general artificial intelligence will not be realized. *Humanities and
  Social Sciences Communications, 7*(1), 1–9.
  https://doi.org/10.1057/s41599-020-0494-4

Floridi, L. (2024). Why the AI hype is another tech bubble. *Philosophy & Technology, 37*(4),
  1–13. https://doi.org/10.1007/s13347-024-00817-w

Future of Life Institute. (2023, March 22). *Pause Giant AI Experiments: An open letter*.
  Future of Life Institute. https://futureoflife.org/open-letter/pause-giant-ai-experiments/

Garud, R., Snihur, Y., Thomas, L. D. W., & Phillips, N. (2023). The dark side of
  entrepreneurial framing: A process model of deception and legitimacy loss. *Academy
  of Management Review*. Advance online publication.
  https://doi.org/10.5465/amr.2022.0213

Gebru, T., & Torres, É. P. (2024). The TESCREAL bundle: Eugenics and the promise of
  utopia through artificial general intelligence. *First Monday, 29*(1).
  https://doi.org/10.5210/fm.v29i1.13568

Goertzel, B., & Pennachin, C. (Eds.). (2007). *Artificial general intelligence*. Springer.

Goldfarb, B., & Kirsch, D. A. (2019). *Bubbles and crashes: The boom and bust of
  technological innovation*. Stanford University Press.

Golumbia, D. (2024). *Cyberlibertarianism: The right-wing politics of digital technology*.
  University of Minnesota Press.

Guston, D. H. (2008). Preface. In E. Fisher, C. Selin, & J. M. Wetmore (Eds.), *The yearbook
  of nanotechnology in society: Presenting futures* (Vol. 1, pp. v–viii). Springer.

Hendrix, J. (2023, May 16). *Transcript: Senate Judiciary Subcommittee hearing on oversight
  of AI* [Transcript]. TechPolicy Press.
  https://www.techpolicy.press/transcript-senate-judiciary-subcommittee-hearing-on-over
  sight-of-ai/

Hu, K., Bajwa, A., Soni, A., & Tong, A. (2025, May 5). OpenAI to remain under non-profit
  control in change to restructuring plans. *Reuters*.
  https://www.reuters.com/business/openai-remain-under-non-profit-control-change-restr
  ucturing-plans-2025-05-05/

Intemann, K. (2022). Understanding the problem of "hype": Exaggeration, values, and trust in science. *Canadian Journal of Philosophy, 52*(3), 279–294. https://doi.org/10.1017/can.2021.30

Jaeger, J. (2023). Artificial intelligence is algorithmic mimicry: Why artificial 'agents' are not (and won't be) proper agents. *History and Philosophy of the Life Sciences, 45*(3), 32. https://doi.org/10.51628/001c.94404

Jasanoff, S., & Kim, S.-H. (2015). *Dreamscapes of modernity: Sociotechnical imaginaries and the fabrication of power*. University of Chicago Press.

Kaplan, J., McCandlish, S., Henighan, T., Brown, T. B., Chess, B., Child, R., Gray, S., Radford, A., Wu, J., & Amodei, D. (2020). *Scaling laws for neural language models*. arXiv. https://arxiv.org/abs/2001.08361

Klingebiel, J. (2022). *A field guide to hype*. https://johannesklingebiel.de/portfolio/a-field-guide-to-hype.html

Knorr-Cetina, K. (1994). Primitive classification and postmodernity: Towards a sociological notion of fiction. *Theory, Culture & Society, 11*(3), 1–22. https://doi.org/10.1177/026327694011003001

Komporozos-Athanasiou, A. (2022). *Speculative communities: Living with uncertainty in a financialized world*. University of Chicago Press.

Kurzweil, R. (2005). *The singularity is near: When humans transcend biology*. Viking.

Landgrebe, J., & Smith, B. (2019). There is no artificial general intelligence. *arXiv*. https://doi.org/10.48550/arXiv.1906.05833

Latour, B. (2013). *An inquiry into modes of existence: An anthropology of the moderns*. Harvard University Press.

Legg S (2011) *Q&A with Shane Legg on risks from AI*. LessWrong. https://www.lesswrong.com/posts/No5JpRCHzBrWA4jmS/q-and-a-with-shane-legg-on-risks-from-ai. Accessed 14 May 2025

Levy, S. (2025, June 4). *Demis Hassabis embraces the future of work in the age of AI*. *Wired*. https://www.wired.com/story/google-deepminds-ceo-demis-hassabis-thinks-ai-will-make-humans-less-selfish/

Lowitzsch, J., & Magalhães, R. (2025). Automation, artificial intelligence and capital concentration: A race for the machine. *International Review of Applied Economics, 39*(2–3), 197–215.

MacAskill, W. (2022). *What we owe the future*. Basic Books.

MacKenzie, D. (2008). *An engine, not a camera: How financial models shape markets*. MIT Press.

Markelius, A., Wright, C., Kuiper, J., Delille, N., & Kuo, Y. T. (2024). The mechanisms of AI hype and its planetary and social costs. *AI and Ethics, 4*(3), 727-742.

Merchant, B. (2025). *AI-generated business: The rise of AGI and the rush to find a working revenue model*. AI Now Institute. https://ainowinstitute.org/publications/ai-generated-business

Millar, N., Batalo, B., & Budgell, B. (2022). Trends in the use of promotional language (hype) in abstracts of successful National Institutes of Health grant applications, 1985–2020. *JAMA Network Open, 5*(8), e2228676. https://doi.org/10.1001/jamanetworkopen.2022.28676

Mitchell, M. (2024). Debates on the nature of artificial general intelligence. *Science, 383*(6689), eado7069. https://doi.org/10.1126/science.ado7069

Morris, M. R., Sohl-Dickstein, J., Fiedel, N., Warkentin, T., Dafoe, A., Faust, A., Farabet, C., & Legg, S. (2023). *Levels of AGI: Operationalizing progress on the path to AGI*. arXiv. https://arxiv.org/abs/2311.02462

Nowotny, H. (2015). *The cunning of uncertainty*. John Wiley & Sons.

OpenAI (2018) *OpenAI charter*. https://openai.com/charter/. Accessed 17 December 2024

OpenAI. (2023, February 24). *Planning for AGI and beyond*. https://openai.com/index/planning-for-agi-and-beyond/

OpenAI. (2025, April 7). *OpenAI's EU economic blueprint*. OpenAI. https://cdn.openai.com/global-affairs/2dbdb523-1a9f-4552-971d-e0948ae2abca/openai-eu-economic-blueprint-apr-2025.pdf

Oreskes, N., & Conway, E. M. (2010). *Merchants of doubt: How a handful of scientists obscured the truth on issues from tobacco smoke to climate change*. Bloomsbury Press.

Paolo G, Gonzalez-Billandon J, Kégl B (2024) *Position: a call for embodied AI*. In: Salakhutdinov R, Kolter Z, Heller K, Weller A, Oliver N, Scarlett J, Berkenkamp F (eds) Proceedings of the 41st International Conference on Machine Learning, vol 235 of Proceedings of Machine Learning Research, pp 39493–39508. PMLR

Pasquale, F. (2020). *New laws of robotics: Defending human expertise in the age of AI*. Harvard University Press.

Pemmaraju, S. (2015). Hedge/hog: Speculative action in financial markets. In V. Rao, P. Krishnamurthy, & C. Kuoni (Eds.), *Speculation, now: Essays and artwork* (pp. 52–59). Duke University Press.

Pfaffenberger, B. (1992). Technological dramas. *Science, Technology, & Human Values*, *17*(3), 282-312.

Pillay, T. (2025, January 8). *How OpenAI's Sam Altman is thinking about AGI and superintelligence in 2025*. *Time*. https://time.com/7205596/sam-altman-superintelligence-agi/

Powers, D. (2012). Notes on hype. *International Journal of Communication, 6*, 17–34. https://ijoc.org/index.php/ijoc/article/view/1224

Schlereth, M. M. (2025). The infinite choice barrier: A triangulated mathematical proof of the impossibility of AGI. *PhilArchive*. https://philarchive.org/rec/SCHTIC-4

Smitsman, A., Goertzel, B., Bozesan, M., & George, L. (2024). Participatory framework for creating a global AGI constitution. *Cadmus, 3*(2024). https://www.cadmusjournal.org/article/volume-3/issue-5/participatory-framework-creating-global-agi-constitution

Späth, J. (2025, March). *The Rise of Techno-Authoritarianism and its Impact on U.S. Foreign Policy* [Policy brief]. Austrian Institute for International Affairs (oiip). https://www.oiip.ac.at/en/publications/the-rise-of-techno-authoritarianism-and-its-impact-on-us-foreign-policy/

Stilgoe, J. (2020). *Who's driving innovation? New technologies and the collaborative state*. Palgrave Macmillan.

Sublime, J. (2024). The AI race: Why current neural network-based architectures are a poor basis for artificial general intelligence. *Journal of Artificial Intelligence Research, 79*, 41–67.

Suleyman, M., & Bhaskar, M. (2023). *The coming wave*. Crown.

Summerfield, C. (2023). *Natural general intelligence: How understanding the brain can help us build AI*. Oxford University Press.

Sun Tzu. (2002). *The art of war* (L. Giles, Trans.). Dover Publications. (Original work published 5th century BCE)

Syme, P. (2023, April 2). A god-like AI needs a safety net, AI investor warns, as heated competition raises the stakes. *Business Insider*. https://www.businessinsider.com/god-like-ai-needs-safety-net-heated-competition-ai-investor-2023-4

Thais S (2024) *Misrepresented technological solutions in imagined futures: the origins and dangers of AI hype in the research community*. In: Proceedings of the AAAI/ACM Conference on AI, Ethics, and Society 7:1455–1465

Torres, P. (2022). *Human extinction: A history of the science and ethics of annihilation*. Routledge.

United Nations. (2024). *Governing AI for humanity: Final report*. United Nations. https://www.un.org/sites/un2.un.org/files/governing_ai_for_humanity_final_report_en.pdf

van Lente H (1993) *Promising technology: the dynamics of expectations in technological developments*. PhD dissertation, Universiteit Twente. https://research.utwente.nl/en/publications/promising-technology-the-dynamics-of-expectations-in-technologica. Accessed 23 January 2025

van Lente, H., Spitters, C., & Peine, A. (2013). Comparing technological hype cycles: Towards a theory. *Technological Forecasting and Social Change, 80*(8), 1615–1628. https://doi.org/10.1016/j.techfore.2012.12.004

Wadhwani, R. D., & Lubinski, C. (2025). Hype: Marker and maker of entrepreneurial culture. *Journal of Business Venturing*, *40*(2), 106455.

Webb, E. (2025, April 28). *'Godfather of AI' says he's 'glad' to be 77 because the tech probably won't take over the world in his lifetime. Business Insider*. https://www.businessinsider.com/ai-godfather-geoffrey-hinton-superintelligence-risk-takeover-2025-4

Wiggers, K. (2024, December 26). Microsoft and OpenAI have a financial definition of AGI, report. *TechCrunch*. https://techcrunch.com/2024/12/26/microsoft-and-openai-have-a-financial-definition-of-agi-report/

Wilkie, A., Savransky, M., & Rosengarten, M. (Eds.). (2017). *Speculative research: The lure of possible futures*. Taylor & Francis.

Winner, L. (1997). Cyberlibertarian myths and the prospects for community. *ACM Sigcas Computers and Society*, *27*(3), 14-19.

Worldcoin (2023) *Worldcoin: a new identity and financial network*. https://whitepaper.worldcoin.org/. Accessed 23 August 2024

Yudkowsky, E. (2023, March 29). *Pausing AI developments isn't enough. We need to shut it all down. Time*. https://time.com/6266923/ai-eliezer-yudkowsky-open-letter-not-enough/

Yudkowsky, E. (2023, May 8). *Eliezer Yudkowsky on the dangers of AI* [Audio podcast episode]. In R. Roberts (Host), *EconTalk*. Dexa.ai. https://dexa.ai/econtalk/d/1282e0ca-8e41-11ee-a86c-573a8fd42009